\documentclass[12pt]{article}
\usepackage{epsfig}
\usepackage{graphicx}
\usepackage{amsmath}
\usepackage{natbib}
\usepackage{geometry}
\newcommand{\be}{\begin{equation}}
\newcommand{\ee}{\end{equation}}
\newcommand{\ba}{\begin{eqnarray}}
\newcommand{\ea}{\end{eqnarray}}

\begin{document}
\bibliographystyle{plain}
\title{\bf Intracellular delay limits cyclic changes\\
 in gene expression}
\author{Katja Rateitschak\footnote{Corresponding author. Tel: +49 381 498 7575, Fax: +49 381 7572}, Olaf Wolkenhauer \\[2ex]
{\small Systems Biology and Bioinformatics Group} \\
{\small University of Rostock} \\
{\small 18051 Rostock, Germany}\\
{\small katja.rateitschak@uni-rostock.de}\\
{\small olaf.wolkenhauer@uni-rostock.de}\\[2ex]
{\small www.sbi.uni-rostock.de}}

\date{\small \today}

\maketitle

\abstract{Based on previously published experimental observations
and mathematical models for Hes1, p53 and NF--$\kappa$B gene
expression,
we improve these models through a distributed delay formulation of
the time lag between transcription factor binding and mRNA production.
This description of natural variability for delays introduces
a transition from a stable steady state to limit cycle oscillations and then
a second transition back to a stable steady state which has not been
observed in previously published models. On the basis of our results and
following recent discussions about the role of delay-induced oscillations in
gene transcription we establish the hypothesis
that the period of the delay-induced cyclic
changes should be characterized by an upper bound so that it cannot
be greater then the period of fundamental biological cycles.
We demonstrate our
approach for two models. The first model
describes Hes1 autorepression with equations for Hes1 mRNA
production and Hes1 protein translation. The second model
describes Hes1 repression by the protein complex Gro/TLE1/Hes1,
where Gro/TLE1 is activated by Hes1 phosphorylation. }

\vspace{1.cm}
\newpage

\section*{\normalsize Introduction}
The three proteins Hes1, p53, and NF--$\kappa$B are
transcriptionally regulated by short negative feedback loops and
actual experiments have revealed that their genes show an
oscillatory expression
\cite{bar-or00,hirata02,hoffmann02,lahav04,nelson04}. In the
vertebrate segmentation clock several cycling genes are involved
in an oscillatory mechanism driving somite segmentation
\cite{giudicelli04}. Mathematical models have been established for
the regulation of these proteins
\cite{bar-or00,hirata02,hoffmann02,nelson04,monk03,lewis03,jensen03,bernard05,ciliberto05,zeiser05}
where the authors of \cite{monk03,lewis03,jensen03,bernard05} have
introduced a discrete time delay to describe the time lag between
transcription factor binding and gene transcription. Analytical
and numerical results have shown that these models, which display
damped oscillations for small positive delay times, can pass a
Hopf bifurcation at a critical delay time where the stable steady
state becomes unstable and a stable limit cycle oscillator emerges
\cite{monk03,lewis03,jensen03,bernard05}.

Naturally, the question arises what the
role of such oscillations in gene transcription could be \cite{lahav04b,nelson04b,barken05,nelson05}?
Due to the fact that the oscillation period is proportional to the
delay time it may be possible that oscillations encode information.
Single cell experiments have shown that the dynamics
of NF--$\kappa$B dependent transcription correlates with the duration of
p65 oscillations \cite{nelson04,nelson04b}. In other single cell experiments p53 has
been expressed in a series of discrete pulses in response to DNA
damage and the average number of pulses over all cells increased
with the damage \cite{lahav04}.

We introduce in our work a new model for gene transcription
networks with a negative feedback loop where the time lag between
transcription factor binding and mRNA production is described, in
line with the natural variability, by a distributed time delay.
With an increasing mean delay time this model can show a Hopf
bifurcation leading to a transition from a stable steady state to
an unstable steady state surrounded by a stable limit cycle
oscillator and then a reverse Hopf bifurcation leading to a
transition from a stable limit cycle oscillator to a stable steady
state. Such a property has not been observed in previously
published mathematical models for gene transcription networks.
Following recent discussions about information encoding in
delay-induced oscillations our results could be interpreted that
the period of the delay-induced cyclic changes cannot be greater
then the period of fundamental biological oscillations, say the
cell cycle or the circadian clock. Therefore for a large delay
time a reverse Hopf bifurcation should lead back to a stable
steady state. This is demonstrated in our model.

The outline of the paper is as follows. The first section introduces
differential equations with discrete and distributed time delays.
In our model the distributed time delay is defined by a kernel of the Gamma function.
The following two sections validate the approach with two models from the literature.
The first model describes Hes1 autorepression with equations for
Hes1 mRNA production and Hes1 protein translation.
The second model describes Hes1 repression by the protein complex Gro/TLE1/Hes1,
where Gro/TLE1 is activated by Hes1 phosphorylation.
Our analytical and numerical results are discussed in relation to experimental data.
This leads us to an investigation into thresholds for damped and sustained oscillations.

\section*{\normalsize Discrete and distributed time delays}
An appropriate mathematical representation to describe a time lag
between an action and a reaction are delay differential equations
\cite{hale93}. If a variable influences the present state of the
system at a fixed time point in the past, $t-\tau$, where $\tau$
is the delay time, then the time lag can be modelled as a discrete
delay
\be
\frac{dx}{dt}=F\left(x,x(t-\tau)\right)\,. \nonumber
\ee
A more realistic description of time lag effects in biological systems
is to assume that the past influences the present state over an
interval of time. Natural variability is weighted by a
probability density $\varrho$, leading to a distributed delay

\be
\frac{dx}{dt}=F\left(x,\int_0^\infty x(t-\tau)\varrho(\tau)d\tau\right)\,.
\label{dd}
\ee

Analyzing a discrete delay differential equation through a Taylor
series expansion of $x(t-\tau)$ leads to an infinite dimensional
system of ordinary differential equations. Approximations by a
finite number of ordinary differential equations have been studied
in \cite{macdonald89,matvii04,mocek05}. In case of a distributed
delay, we discuss below an example where a finite set of ordinary
differential equations can exactly be derived. In either case, the
increased dimension can lead to a stabilization or destabilization
of the dynamics in  feedback models.

Delay differential equations have been extensively applied to biological
and engineering processes.
This includes gene transcription
\cite{monk03,lewis03,jensen03,bernard05,smolen98,smolen99};
the nucleo--cytoplasmic translocation of proteins in eucaryotes \cite{swameye03,nikolov05};
 the hatching period and the incubation time in population models \cite{may01,eurich05}.
In control theory the delay between the observation and the
control of a variable has been considered  \cite{franklin02,just03}.

A frequently used model for distributed time delays in biological
applications is to choose for the probability density $\varrho$ in
Eq.~(\ref{dd}) the kernel of the Gamma function \cite{macdonald89}
\be
g_q^p(\tau)=\frac{q^p}{(p-1)!}\tau^{p-1} e^{-q\tau}
\label{gamma}\,,
\ee
which has a maximum at $\tau=p/q$
and is zero at $\tau=0$ and for $\tau\to\infty$. The mean delay
time is $\bar{\tau}=(p+1)/q$.
This leads us to the following model
\be
\frac{dx}{dt}=F\left(x,\int_0^\infty x(t-\tau)g_q^p(\tau)d\tau\right) \nonumber
\ee
This model has
the advantage that using the linear chain trick \cite{macdonald89} we can transform the delay
differential equations into a finite number of ordinary
differential equations.

Authors that used this idea have found that distributed delay differential
equations lead to a larger range of stability than discrete delay differential equations
\cite{eurich05,bernard01,thiel03}. This is important in ecological models where the destabilization of the dynamics
can lead to an extinction of species \cite{eurich05} or in physiological models where the disease is
characterized by a stability change of a variable \cite{bernard01}.

\section*{\normalsize Hes1 autorepression (Model I)}
A mathematical model for Hes1 protein autorepression with a discrete time delay has
been introduced in \cite{monk03,lewis03,jensen03,bernard05}. The discrete time delay describes the time lag between
Hes1 protein binding to the regulatory DNA and the production of Hes1 mRNA.
For a critical delay time this model passes a Hopf bifurcation where the stable steady state is left and the system
moves to a stable limit cycle \cite{monk03,lewis03,jensen03,bernard05}.
We here modify this model by replacing in Eqs.~(3.1) and (3.2) of
Ref.~\cite{bernard05} the discrete time delay by a distributed time delay with Gamma kernel of Eq.~\eqref{gamma}, leading to
\be
\left.
\begin{aligned}
\frac{d}{dt}mRNA &=\frac{b\cdot k^h}{k^h+(\int_0^\infty Hes\mathit{1}(t-\tau)\cdot g_q^p(\tau)d\tau)^h}-a\cdot mRNA \\
\frac{d}{dt}Hes\mathit{1} &= b\cdot mRNA-a\cdot Hes\mathit{1}\,.
\label{A-model}
\end{aligned}
\quad\right.
\ee

The first equation describes the temporal change of the Hes1 mRNA
concentration, which depends on the autorepression by the Hes1
protein. The second equation describes the translation of the
mRNA.
We have implemented the distributed delay such that the averaging is performed
over the transcription factor concentration reflecting the varying binding to
their sites and the dissociation.
In the linear case the smaller translation delay can be shifted into the transcription
term without changing the dynamics \cite{monk03}.
Without loss of generality, we simplify Eqs.~(3.1)
and (3.2) of Ref.~\cite{bernard05} by choosing the same rate constants
for transcription and translation and also the same rate
constant for degradations (cf.~Table
\ref{tab-res}). We have chosen the parameter for the degradation
rates ten-fold higher than in \cite{bernard05} so that the delay
times which induce limit cycle oscillations are in the range of the
temporal duration of gene transcription (10--40 min).

The first step in analyzing the properties of differential equations
is to look for steady states and to study their stability.
 For
$k^h=(b/a)^h\cdot a/(b-a)$ the steady state of Eqs.~(\ref{A-model}) is $mRNA^*=1$ and $Hes\mathit{1}^* =b/a$ which is stable
for the case without delay \cite{bernard05}.
Linear stability analysis of Eqs.~(\ref{A-model}) at steady state leads to the following eigenvalue equation

\be
(\lambda+a)^2+\frac{a^2h}{b}(b-a)\int_0^\infty g_q^p(\tau)\cdot e^{-\lambda\tau}d\tau=0 \,. \nonumber
\ee
The integral can be
analytically solved, resulting in
\be
(\lambda+a)^2+\frac{a^2h}{b}(b-a)\frac{q^p}{(\lambda+q)^p}=0\,.\nonumber
\ee

The eigenvalue equation is independent of how the averaging of
the distributed time delay in Eqs.~(\ref{A-model}) is performed.
The eigenvalue equation is the same regardless whether $h$ is
outside the integral as in Eqs.~(\ref{A-model}) or whether $h$ is an exponent of $Hes\mathit{1}(t-\tau)$.
The eigenvalue equation is also the same if we average over the transcription rate, i.~e.~,
\be
\int_0^\infty\frac{b\cdot k^h}{k^h+Hes\mathit{1}(t-\tau)^h}\cdot g_q^p(\tau)d\tau\,.\nonumber
\ee

But
it is different to the eigenvalue equation for the related model
with a discrete time delay, Eq.~(A13) in \cite{bernard05}. Thus
considering natural variability in a delay differential equation
model for Hes-protein autorepression can lead to different
steady state dynamics, but the steady state dynamics is
independent of the source of natural variability.

\begin{figure}
\centerline{\epsfxsize=25cm\epsfbox{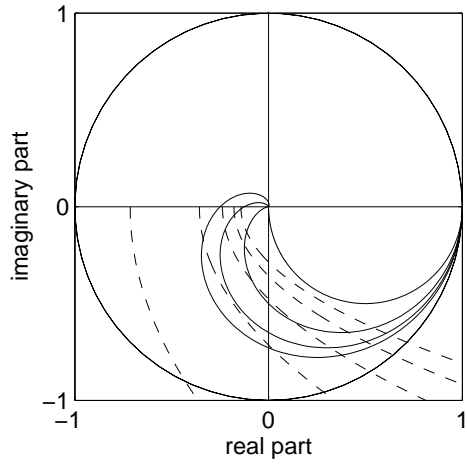}}
\caption{Model I: Intersections between the delay curves (solid
lines) and the ratios curves (dashed lines). According to
Eq.~(\ref{A-evtrafo}) intersections indicate putative Hopf
bifurcations. The delay curves of the distributed time delay are
shown for $p=1$ (in first quadrant only), 2, 3, 4 and the ratio
curves are shown for $h=2$ (leftmost curve), 4, 6, 8, 10. In
addition the unit circle as the discrete time delay curve is
shown.} \label{A-graphmeth}
\end{figure}

 The next step is to analyze how the delay time influences the stability of the steady state by identifying critical mean
delay times $\bar\tau=(p+1)/q$, which form Hopf bifurcation points. First we apply the graphical method from
Ref.~\cite{macdonald89} to get an immediate overview of putative Hopf bifurcations. We transform the eigenvalue equation
with $\lambda=\sigma+i\omega$
and $\sigma=0$ to

\be
-\frac{b(i\omega+a)^2}{a^2h(b-a)}=\frac{q^p}{(i\omega+q)^p}
\label{A-evtrafo}
\ee
Now the delay is separated into the delay curve on the right--hand--side (r.h.s.) of Eq.~(\ref{A-evtrafo})
and the ratio curve on the left--hand--side (l.h.s.). Both curves are plotted as a function of $\omega$ in the complex plane
 in Fig.~\ref{A-graphmeth}, together with the unit circle as the discrete delay curve.
The shape of the delay curve does not depend on the scaling factor $q$.
Intersections between a delay curve and a ratio curve are putative
Hopf bifurcations. For several parameter pairs $(p,h)$ two intersections occur indicating putative
steady state -- limit cycle -- steady state transitions.
For low values of $p$ the system remains stable for low $h$ or can pass through a
steady state -- limit cycle -- steady state transition for higher $h$.
For intermediate values of $p$ three scenarios are possible: stable steady state for low $h$,
steady state -- limit cycle -- steady state transitions for intermediate $h$ and
steady state -- limit cycle transition for high $h$.
For $p\to\infty$ the delay curve converges to the discrete delay curve.

For $p=1$ and $p=2$,  Eq.~(\ref{A-evtrafo}) can be analytically solved for $q$

\be
q_{\pm}=a\left(c_p-1\pm\sqrt{(c_p-1)^2-1}\right)
\label{sol-A}
\ee
with
\be
c_{p=1}=\frac{h}{4b}(b-a) \quad\mbox{and}\quad
c_{p=2}=\frac{h}{2b}(b-a) \nonumber \,.
\ee
The critical mean delay
times $\bar\tau_{min}$ and $\bar\tau_{max}$ for stability changes
can be calculated from $q_{\pm}$
\be
\bar\tau_{min}=\frac{p+1}{q_+} \quad\mbox{and}\quad
\bar\tau_{max}=\frac{p+1}{q_-} \nonumber\,.
\ee
The parameter $q$ in Eqs.~(\ref{A-model}) is chosen as the bifurcation parameter related to $\bar\tau$.
The results for $\bar\tau_{min}$ and $\bar\tau_{max}$ are summarized in Table \ref{tab-res}.
This is compared to the results for the discrete delay according to Eq.~(3.4) of
\cite{bernard05}.
The delay times of the two putative Hopf bifurcation points $\bar\tau_{min}$ and $\bar\tau_{max}$
are in the range of the duration of gene transcription.
In addition, the results for $a=0.03$ (parameter value in \cite{monk03,bernard05}) are shown in Table \ref{tab-res2}.
For $a=0.03$ only the delay time of the first Hopf bifurcation point $\bar\tau_{min}$ is in the range of the
duration of gene transcription.

Comparing the distributed delay with the discrete delay shows
that the distributed delay destabilizes at a higher critical mean delay time than the discrete delay.
This is in agreement with the results in Refs.~\cite{eurich05,bernard01,thiel03}, where the authors have found that the
distributed delay leads to a larger range of stability than with a discrete delay.

The following last step in our analysis is to check whether $\sigma$ changes its sign at the critical values
$\bar\tau_{min}$ and $\bar\tau_{max}$. For most cases this is difficult or impossible to calculate analytically,
 which is why we used numerical simulations to check whether the two predicted Hopf
bifurcations are passed.
We transform Eqs.~(\ref{A-model}) with $p=2$ using the linear chain trick \cite{macdonald89} to obtain
\be
\left.
\begin{aligned}
\frac{d}{dt}x_0&=q\cdot(Hes\mathit{1}-x_0)\\
\frac{d}{dt}x_1&=q\cdot(x_0-x_1)\\
\frac{d}{dt}mRNA &=\frac{b\cdot k^h}{k^h+x_1^h}-a\cdot mRNA \\
\frac{d}{dt}Hes\mathit{1} &= b\cdot mRNA-a\cdot Hes\mathit{1} \,.
\label{A-ode}
\end{aligned}
\quad\right.
\ee

\begin{figure}
\centerline{\epsfxsize=20cm\epsfbox{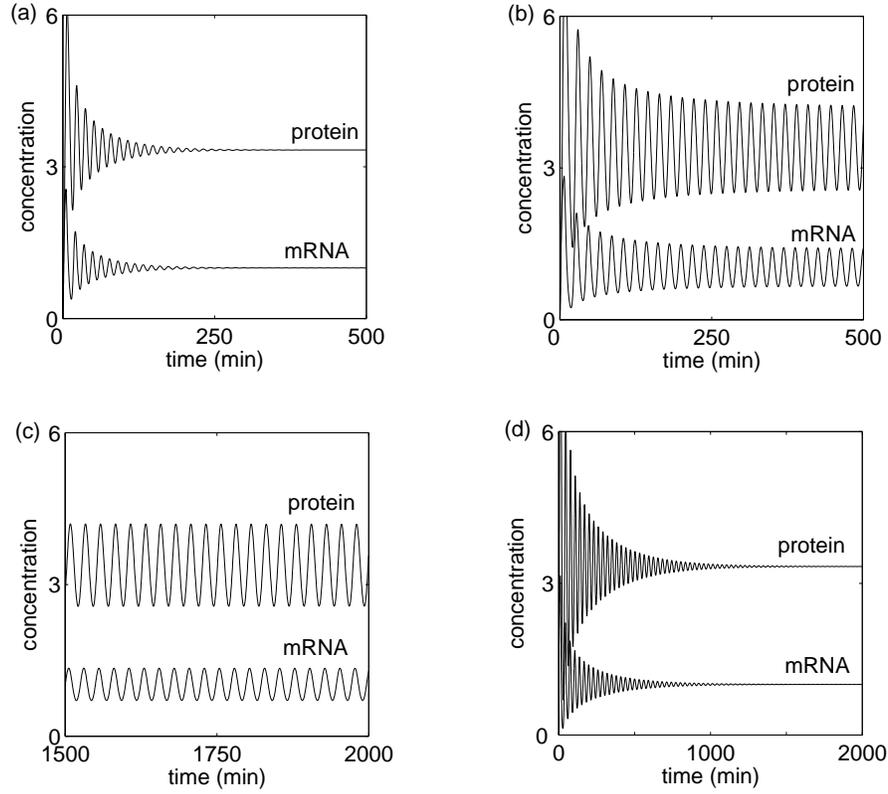}}
\caption{Model I: Numerical simulations of the exactly derived
ordinary differential equations Eqs.~(\ref{A-ode}) with $p=2$ and
$h=6$, showing a steady state -- limit cycle -- steady state
transition by passing two Hopf bifurcation points with increasing
mean delay time. (a) $\bar\tau=4$, (b) $\bar\tau=8$, (c)
$\bar\tau=14$, (d) $\bar\tau=20$. Beyond the reverse Hopf
bifurcation the damping of the oscillations increases with the
delay time. For $\bar\tau=50$ the steady state is already reached
at a time of $300$min.} \label{A-simu}
\end{figure}

Equations (\ref{A-ode}) represent a monotone cyclic feedback system.
Several authors have focused on the existence of periodic solutions
in such kind of equations \cite{goodwin65,tyson78,mahaffy80,mallet-paret90}.
According to the Poincar\'{e}--Bendixson theorem
for monotone cyclic feedback systems a stable
solution of Eqs.~(\ref{A-ode}) is either a stable steady state or
a stable limit cycle oscillator \cite{mallet-paret90}. Results from numerical
simulations of Eqs.~(\ref{A-ode}) with $h=6$ are presented in
Fig.~\ref{A-simu}. The time course of the mRNA and Hes1 protein
concentrations for delay times below and above the critical mean
delay times show that the bifurcation at $\bar\tau_{min}$
leads to a transition from a stable steady state to an unstable
steady state. The latter is surrounded by a stable limit cycle oscillator.
The reverse bifurcation at $\bar\tau_{max}$ leads back to a stable steady state.
Table \ref{tab-freq-amp} shows the frequency and the amplitude of the limit cycle oscillations as a function
of the mean delay time.
The frequency is inverse proportional to the delay time while the amplitude has a maximum between the bifurcation points.
Information can only be encoded in the frequency of the oscillations.

It follows that our model of Hes1 autorepression with a
distributed delay has a maximal delay time for delay--induced
limit cycle oscillations. This result is different to the discrete
delay models \cite{monk03,lewis03,jensen03,bernard05} which either
show a steady state -- limit cycle transition or no bifurcation.
This can straightforwardly be shown for a generalized eigenvalue
equation for these models, which reads
$(A-\lambda)\cdot(B-\lambda)-C\cdot e^{-\lambda\tau}=0$. The
solution for purely imaginary eigenvalues $\lambda=i\omega$ is $
w^2=-(A^2+B^2)/2\pm\sqrt{((A^2+B^2)^2/4-A^2B^2+C^2)}$ and
$\tau=1/\omega\cdot\arccos((AB-\omega^2)/C)$. Multiple solutions
for $\tau$  due to the periodicity of the angular function do not
change stability \cite{macdonald89}. Thus these two-dimensional
negative feedback models for gene transcription with discrete time
delay either show no Hopf bifurcation or they can show only one
Hopf bifurcation but they can never show a Hopf bifurcation and a
reverse Hopf bifurcation with increasing delay time. In contrast,
a damped harmonic oscillator as an example from physics modelled
with a discrete time delay can also show alternating stability and
instability with increasing delay time \cite{macdonald89}.

The steady state properties of our model are independent on the
way the averaging is performed in the distributed time delay. But
they can depend on the kernel of the delay distribution. An
alternative approach for a distributed time delay has been studied
in \cite{monk03}. The averaging is performed over the
transcription rate but with a uniform kernel of the distribution
on a finite support. In contrast to our results the results in \cite{monk03} are
indistinguishable from the discrete delay case.
From our point of view, the Gamma kernel is a more
realistic description for the probability density of a distributed time delay,
because it has a maximum and approaches zero for $\tau\to 0$
and $\tau \to \infty$.

The infinite support of the delay distribution in our model does
not lead to an interference of the delay induced oscillations with
fundamental biological oscillations for a broad parameter range
because with increasing delay time the kernel of the Gamma function
exponentially converges to zero.
We calculate a boundary $max$ such that the contribution
of the tail of the delay distribution can be neglected:
\be
\left(\int_0^{max} Hes\mathit{1}(t-\tau)g_q^p(\tau)d\tau+\int_{max}^\infty Hes\mathit{1}(t-\tau)g_q^p(\tau)d\tau\right)^h\,.\nonumber
\ee
We consider only the expression with the highest contribution of the tail in this sum.
Assuming $Hes\mathit{1}=10$ as an upper boundary for the protein concentration.
Lower boundaries for the parameter $max$ for which
the inequality
\be
h\cdot\left(\int_0^{max} Hes\mathit{1}(t-\tau)g_q^p(\tau)d\tau\right)^{h-1}\cdot\int_{max}^\infty Hes\mathit{1}(t-\tau)g_q^p(\tau)d\tau<10^{-6}\nonumber
\ee
holds are $max=176$min for $h=6$,
$\bar\tau=16$min  and $max=672$min for $h=10$,$\bar\tau=48$min (critical
delay times of reverse Hopf bifurcations). They are smaller for
example then a typical time interval between two cell divisions in
mammals of 24 hours or the circadian clock period.

\section*{\normalsize Gro/TLE1 mediated repression of Hes1 (Model II)}
\begin{figure}
\centerline{\epsfxsize=25cm\epsfbox{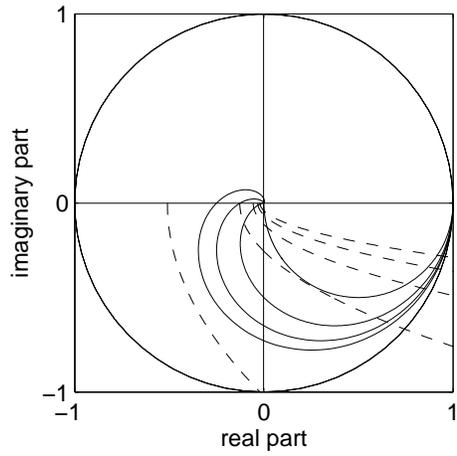}}
\caption{Model II: Intersections between the delay curves (solid
line) and the ratio curves (dashed line). According to
Eq.~(\ref{B-evtrafo}) intersections indicate putative Hopf
bifurcations. The delay curves are shown for $p=1$ (first quadrant
only), 2, 3, 4 and the ratio curves are shown for $h=2$ (most left
curve), 4, 6, 8, 10.} \label{B-graphmeth}
\end{figure}

To see whether other models with a negative feedback loop can show a
steady state -- limit cycle -- steady state transition, we consider what is referred to as Model C in \cite{bernard05}.
Here Hes1 repression is realized by the protein complex Gro/TLE1/Hes1, denoted $GroH$, where Gro/TLE1 is activated by Hes1 phosphorylation.
We replace in Eqs.~(4.1) and (4.2) of Ref.~\cite{bernard05}  the
discrete delay by a distributed delay with the Gamma kernel
\be
\left.
\begin{aligned}
\frac{d}{dt}Hes\mathit{1} &= \frac{b\cdot k^h}{k^h+(\int_0^\infty g_q^p(\tau)\cdot GroH(t-\tau)d\tau)^h}-a\cdot Hes\mathit{1} \\
\frac{d}{dt}GroH &= \frac{b\cdot Hes\mathit{1}^h}{l^h+Hes\mathit{1}^h}-a\cdot GroH \,.
\label{B-model}
\end{aligned}
\quad\right.
\ee
For $k^h=a/(b-a)$ and $l^h=(b-a)/a$ the steady
state is $Hes\mathit{1}^*=1$ and $GroH^*=1$. Linear stability
analysis leads to the following eigenvalue equation at the steady
state
\be
0=(\lambda+a)^2+\frac{a^2h^2}{b^2}(b-a)^2\frac{q^p}{(\lambda+q)^p}\nonumber
\ee
Again we apply the graphical method described
above to get an overview of putative Hopf bifurcations or reverse
Hopf bifurcations. Transformation of the eigenvalue equation with
$\lambda=\sigma+i\omega$ and $\sigma=0$ leads to
\be
-\frac{b^2(i\omega+a)^2}{a^2h^2(b-a)^2}=\frac{q^p}{(i\omega+q)^p}
\label{B-evtrafo}
\ee
The ratio curve (l.h.s.) and the delay curve
(r.h.s.) of Eq.~(\ref{B-evtrafo}) are shown in
Figure~\ref{B-graphmeth}. Several parameter pairs $(p,h)$ show two
putative Hopf bifurcations indicating stable steady state -- limit
cycle -- stable steady state transitions.

The critical parameter $q$ can be analytically calculated from  Eq.~(\ref{B-evtrafo})
for $p=1$ and $p=2$
\be
q_{\pm}=a\left(c_p-1\pm\sqrt{(c_p-1)^2-1}\right) \label{sol-B}
\ee
with
\be
c_{p=1}=\frac{h^2}{4b^2}(b-a)^2 \quad\mbox{and}\quad c_{p=2}=\frac{h^2}{2b^2}(b-a)^2\,. \nonumber
\ee
The results for the critical mean delay times $\bar\tau_{min}$ and $\bar\tau_{max}$ are summarized in Table \ref{tab-res},
together with the result for the discrete delay (according to Eq.~(4.4) in \cite{bernard05}).
The delay times of the two putative Hopf bifurcation points $\bar\tau_{min}$ and $\bar\tau_{max}$
are for some $(p,h)$ pairs close to the range of the duration of gene transcription.
The results for $a=0.03$ are shown in Table \ref{tab-res2}.
For $a=0.03$ only the delay time of the first Hopf bifurcation point $\bar\tau_{min}$ is close to the range of the
duration for gene transcription.

Comparing the results for the distributed delay with those for the discrete delay shows
that the distributed delay destabilizes at a larger critical mean delay time than the discrete delay.
This is in agreement with the results for Model I.
Comparing now the critical delay times of Model I and II shows that Model II destabilizes at a lower critical delay time
$\bar\tau_{min}$ and stabilizes at a higher critical delay time $\bar\tau_{max}$ than Model I.
This is related to the stronger curvature of the ratio
curve of Model II, which comes from cooperative repression together with cooperative activation
whereas in Model I only cooperative repression is included.

\begin{figure}
\centerline{\epsfxsize=20cm\epsfbox{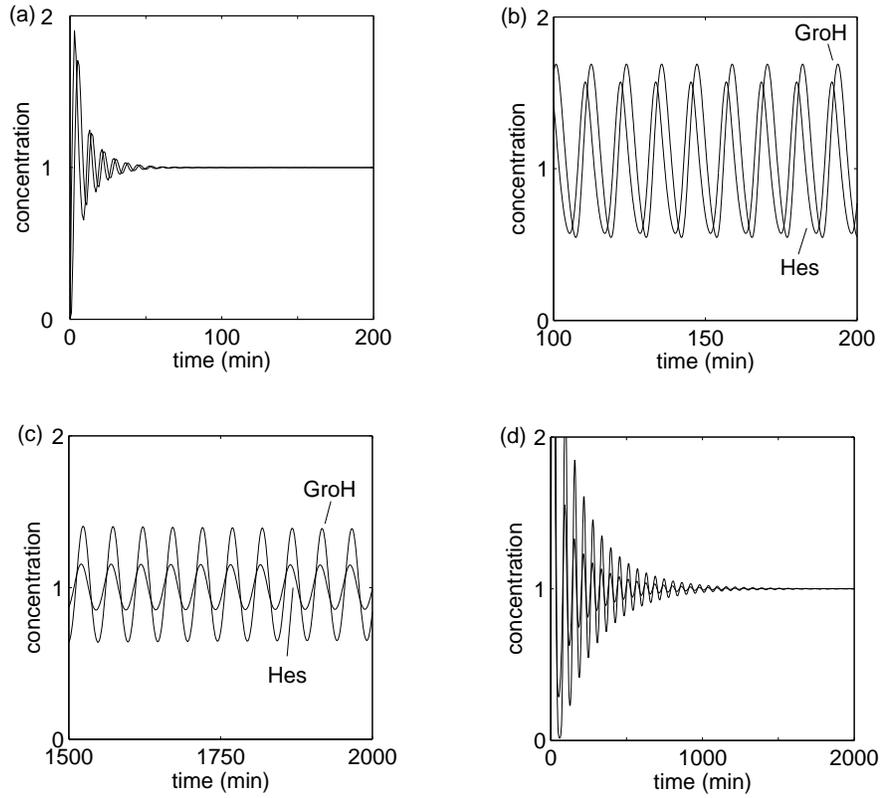}}
\caption{Model II: Numerical simulations of the exactly derived
ordinary differential equations Eqs.~(\ref{B-ode}) with $p=2$ and
$h=4$, showing a steady state -- limit cycle -- steady state
transition by passing two Hopf bifurcation points with increasing
mean delay time. (a) $\bar\tau=1$, (b) $\bar\tau=3$, (c)
$\bar\tau=55$, (d) $\bar\tau=70$.} \label{B-simu}
\end{figure}

For numerical simulations, we use the linear chain trick to derive for  $p=2$ the ordinary differential equations
\be
\left.
\begin{aligned}
\frac{d}{dt}x_0&=q\cdot(GroH-x_0)\\
\frac{d}{dt}x_1&=q\cdot(x_0-x_1)\\
\frac{d}{dt}Hes\mathit{1} &= \frac{b\cdot k^h}{k^h+x_1^h}-a\cdot Hes\mathit{1} \\
\frac{d}{dt}GroH &= \frac{b\cdot Hes\mathit{1}^h}{l^h+Hes\mathit{1}^h}-a\cdot GroH\,,
\label{B-ode}
\end{aligned}
\quad\right.
\ee
The simulations confirm the two Hopf bifurcations  as shown in Figure~\ref{B-simu}.
The bifurcation at $\bar\tau_{min}$
leads to a transition from a stable steady state to an unstable steady state, surrounded by a stable limit cycle
oscillator. The reverse bifurcation at $\bar\tau_{max}$ leads to a transition back to a stable steady state.

\section*{\normalsize Comparison with experimental results}
In signaling and damage repair a fast response is desirable,
which suggests that the initial reaction period is most important.
This means that both, sustained and damped oscillations, are relevant for information encoding.
The authors of Ref.~\cite{lahav04} have measured discrete pulses for protein concentrations with fixed amplitude
and duration, which do not depend on the amount of the stimulus. The experimental results in Ref.~\cite{nelson04}
show only few periods of the oscillations,
making it difficult to decide whether one deals with sustained or damped oscillations.
The same may be said for the results in \cite{hirata02}.
A further complication arises from the difference between single-cell measurements,
compared to averaging measurements.
From a modelling perspective, sustained oscillations that arise
after passing a Hopf bifurcation point, have a clear delay time threshold
for frequency encoding signal transduction.

We now discuss possible thresholds for frequency encoding in dam\-ped oscillations.
The mathematical models of Eqs.~\eqref{A-model} and \eqref{B-model}
show damped oscillations for large ranges of parameters
without passing a Hopf bifurcation at a critical delay time.
For other parameter ranges they approach a constant
steady state. If damped oscillations are indeed relevant then the
question after a threshold arises. A threshold is important
to filter out noise which is an inherent property of biological
signaling pathways. For damped oscillations a threshold could be the
amplitude, which is however more sensitive to noise than the frequency of oscillations.

\begin{figure}
\centerline{\epsfxsize=20cm\epsfbox{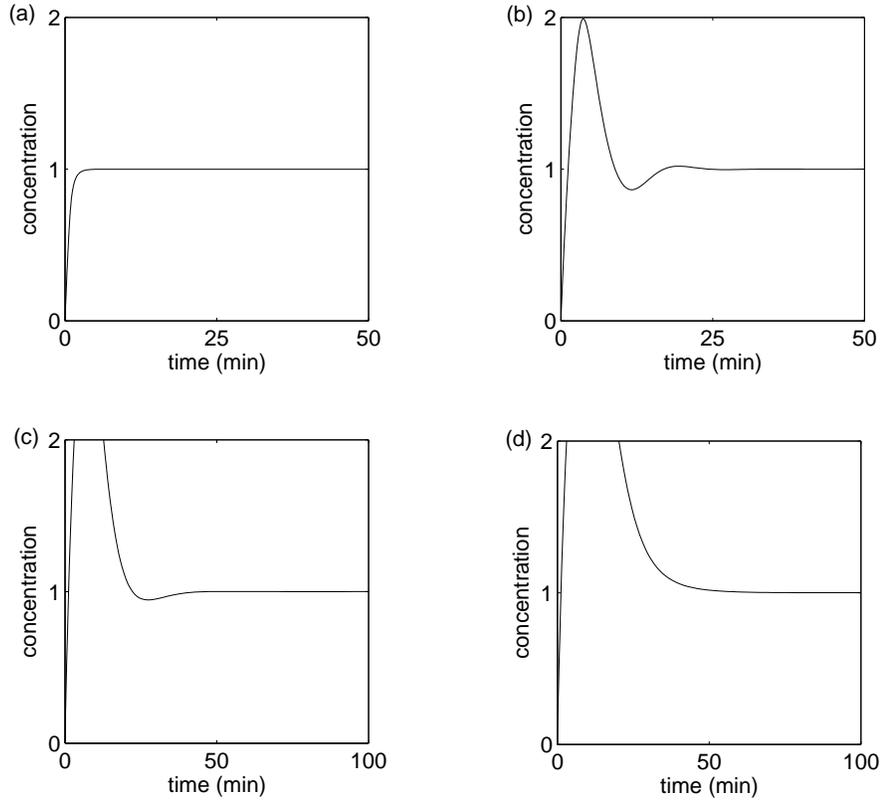}}
\caption{Model III: Numerical simulations of exactly derived
ordinary differential equations from Eq.~(\ref{1d-h}) with $p=1$
and $h=4$, showing no pronounced damped oscillations. (a)
$\bar\tau=0.3$ (b) $\bar\tau=10$ (c) $\bar\tau=50$ (d)
$\bar\tau=100$. Within the delay time interval $[0.5,87]$ the
eigenvalues are complex according to Eq.~(\ref{1d-crit}).}
\label{1-damped}
\end{figure}

\begin{figure}
\centerline{\epsfxsize=20cm\epsfbox{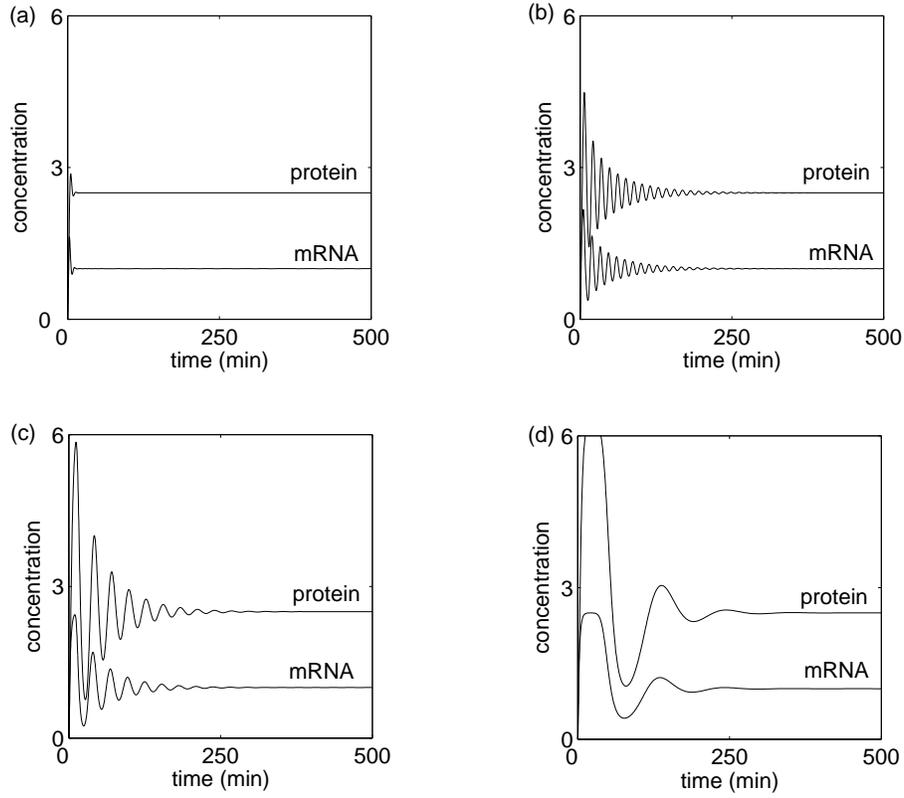}}
\caption{Model I: Numerical simulations of the exactly derived
ordinary differential equations Eqs.~(\ref{A-ode}) with $p=2$,
$h=6$ and $a=0.4$, showing for intermediate delay times pronounced
damped oscillations without passing a Hopf bifurcation. (a)
$\bar\tau=0$, (b) $\bar\tau=5$, (c) $\bar\tau=20$, (d)
$\bar\tau=100$.} \label{A-damped}
\end{figure}

An alternative threshold for damped oscillation could be the critical
delay time at which for the first time at least two eigenvalues switch from real
numbers to complex numbers; while the real parts of all eigenvalues remain negative.
Such a threshold can be calculated analytically for a one-dimensional Hes1 protein autorepression model (Model III)
 \be
 \frac{d}{dt}Hes\mathit{1} = \frac{b\cdot k^h}{k^h+(\int_0^\infty g_q^p(\tau)\cdot Hes\mathit{1}(t-\tau)d\tau)^h}-a\cdot Hes\mathit{1}
\label{1d-h}
 \ee
Linear stability analysis of Eq.~(\ref{1d-h}) with $p=1$ leads to
the following eigenvalue equation
\be
0=\lambda+a+c\cdot\frac{q}{(\lambda+q)} \quad\mbox{with}\quad c=\frac{ha(b-a)}{b} \nonumber
\ee
resulting in
\be
\lambda_{1/2}=-\frac{a+q}{2}\pm\sqrt{\left(\frac{(a+q)^2}{4}-a\cdot q-c\cdot q\right)} \,.\nonumber
\label{1d-ev}
\ee
From the zeros of the term under the square root
\be
q_{1/2}=a+2\cdot c\pm\sqrt{\big((a+2\cdot c)^2-a^2\big)}
\label{1d-crit}
\ee
we calculate the mean delay
times $\bar\tau=(p+1)/q$ for transitions between real and complex
eigenvalues. For $\bar\tau$ below and above the transition values numerical
simulations of exactly derived ordinary differential equations
from Eq.~\eqref{1d-h} show no qualitative difference in
the dynamics of the Hes1 protein concentrations, as shown in Figure
\ref{1-damped}. On the other hand, numerical simulations of Model
I with parameters $h=6$, $p=2$ and $a=0.4$, display for
intermediate delay times damped oscillations with higher natural
frequency, while for no and large delay times the oscillations
have a lower natural frequency, as shown in Figure \ref{A-damped}.
The eigenvalues  related to these parameters are complex for all
delay times $\tau \ge 0$ but no Hopf bifurcation is passed.
The results shown in Figures \ref{1-damped} and \ref{A-damped}
lead us to the conclusion that that the first switch of two eigenvalues from real
numbers to complex numbers is not an appropriate thereshold.

Based on the results shown in Figures \ref{1-damped} and \ref{A-damped}
we suggest as a quantitative measure for information encoding in the frequency
of damped oscillations the ratio of frequency and damping defined by
\be
\zeta=\frac{\omega}{\gamma}\,.\nonumber
\ee
The values for $\omega$ and $\gamma$ are obtained from curve fits for the data in Figures \ref{1-damped} and \ref{A-damped}
using the following equation
\be
y=\alpha\cdot e^{-\gamma t}\cdot \cos(\omega t+\varphi)\,.\nonumber
\ee
This is motivated by the concept of a damping ratio in control engineering \cite{franklin02}.
The results for $\zeta$ are summarized in Table \ref{tab-damping-ratio}.
The highest values for $\zeta$ have the curves in Figure \ref{A-damped} (b) and (c).

\section*{\normalsize Summary and conclusions}
In the present paper we investigated mathematical models for gene
transcription networks with negative feedback loops. We described the
time lag between transcription factor binding and mRNA production
by a distributed time delay compared to previously published
models which use a discrete delay.
We applied our approach to describe experimentally observed
oscillations of the transcription factor Hes1.

We showed that a distributed time delay can lead to
limit cycle oscillations for a finite range of mean
delay times $[\bar\tau_{min},\bar\tau_{max}]$.
We believe this to be a more realistic property of gene transcription
networks with negative feedback if information is encoded in delay--induced oscillations.
More specifically, we note that any oscillation period
should not exceed the period of fundamental cellular processes, e.g.~the cell cycle.
This is an implicit property of our model with a Hopf bifurcation and a reverse Hopf bifurcation
leading to steady state -- limit cycle -- steady state transitions.

Due to the possibility that both, sustained and damped oscillations could have a role in
signal transmission we discussed the encoding of information in the frequency of the
oscillations. Towards this end we proposed a quantitative measure for damped oscillations.

For future experimental work we suggest a more detailed experimental study into the nature of oscillations in gene transcription,
investigating the consequences
of pathological mutations that lead to abnormal time lags between
transcription factor binding and mRNA production.

\subsection*{Acknowledgements}
We thank Nicholas Monk and Volkmar Liebscher for helpful comments on the manuscript.
K.R. acknowledges support by the ministry for education, science and culture of the state Mecklenburg-Vorpommern
and the ``European Regional Development Fund'' (ERDF).
O.W.'s contributions were supported by the European Community as part of the FP6 project COSBICS and by the
German Federal Ministry for Education \& Research (BMBF) as part of the NGFNII SMP Protein.

\newpage


\begin{table}
\begin{center}
\begin{tabular}{|rr|rr|r|}
\hline
\multicolumn{2}{|c|}{Model} &  \multicolumn{2}{|c|}{distributed}  & discrete  \\
&&$\bar{\tau}_{min}$ & $\bar{\tau}_{max}$ & $\tau$ \\
\hline
& & & & \\
\multicolumn{2}{|c|}{I:} & & & \\
$h$ & $p$ & & &\\
6 & 2 & 6 & 16 & 2 \\
8 & 2 &  3 & 33  & 0.2 \\
10 &2 & 2 & 48 & 0.1 \\
\hline
& & & & \\
\multicolumn{2}{|c|}{II:} & & & \\
$h$ & $p$ & & &\\
4 & 2 & 2 & 57 & 0.9\\
6 & 2 & 0.6 & 156 & 0.4 \\
8 & 2 & 0.3 & 293 & 0.2\\
10 & 2 & 0.2 & 470 & 0.1 \\
6 & 1 & 1 & 44 & 0.4\\
8 & 1 & 0.5 & 91 & 0.2 \\
 \hline
\end{tabular}
\end{center}
\caption{Putative Hopf bifurcation points for distributed and discrete time delays. Delay times are given in minutes (min).
For $a=0.3$min$^{-1}$ and
$b=1$min$^{-1}$ analytical results from eigenvalue equation Eq.~(\ref{sol-A}) for Model I
and Eq.~(\ref{sol-B}) for Model II.
We have chosen the degradation rate $a$ ten-fold higher than
in Ref.~\cite{bernard05} to ensure that the delay times are in the range of
the time for gene transcription.}
\label{tab-res}
\end{table}

\begin{center}
\begin{table}
\begin{center}
\begin{tabular}{|rr|rr|r|}
\hline
\multicolumn{2}{|c|}{Model} &  \multicolumn{2}{|c|}{distributed}  & discrete  \\
& & $\bar{\tau}_{min}$ & $\bar{\tau}_{max}$ & $\tau$  \\
\hline
& & & & \\
\multicolumn{2}{|c|}{I:} & & & \\
$h$ & $p$ & & &\\
6 & 2 & 28 & 354 & 13 \\
8 & 2 &  18 & 558  & 9 \\
10 & 2 & 13 & 757 & 7 \\
& & & &\\
10& 1 & 27 & 163 & 7 \\
\hline
& & & & \\
\multicolumn{2}{|c|}{II:} & & & \\
$h$ & $p$ & & & \\
4 & 2 & 8 & 1298 & 5\\
6 & 2 & 3 & 3184 & 2\\
8 & 2 & 2 & 5820 & 1\\
10 & 2 & 1 & 9208 & 1\\
& & & & \\
4 & 1 & 12 & 356 & 5\\
6 & 1 & 4 & 991 &  2 \\
8 & 1 & 2 & 1872 & 1\\
10 & 1 & 1 & 3002 & 1\\
\hline
\end{tabular}
\end{center}
\caption{Putative Hopf bifurcation points, analytical results from
eigenvalue equation Eq.~(\ref{sol-A}) for Model I and  Eq.~(\ref{sol-B})
for Model II with $a=0.03$min$^{-1}$ and $b=1$min$^{-1}$.}
\label{tab-res2}
\end{table}
\end{center}

\begin{table}
\begin{center}
\begin{tabular}{|r|r|r|}
\hline
$\bar\tau$ & frequency & amplitude \\
\hline
& & \\
8 & 0.053 & 1.5 \\
10 & 0.048 & 2.0 \\
12 & 0.043 & 2.0 \\
14 & 0.040 & 1.7 \\
\hline
\end{tabular}
\end{center}
\caption{Model I: Frequency and amplitude of limit cycle oscillations with increasing delay time
for $p=2$, $h=6$ and $a=0.3$min$^{-1}$.}
\label{tab-freq-amp}
\end{table}

\begin{table}
\begin{center}
\begin{tabular}{|r|r||r|r|r|}
\hline
& $\zeta=\omega/\gamma$ & & \multicolumn{2}{|c|}{$\zeta=\omega/\gamma$}  \\
\hline
$\bar\tau$  & III: & $\bar\tau$ & I: mRNA & I: protein \\
\hline
& & & & \\
0.3 & 0 & 0 & 2 & 2 \\
10 & 2 & 5 &  13 & 12 \\
50 & 1 & 20 & 7 & 8 \\
100 & 1 & 100 & 3 & 3 \\
\hline
\end{tabular}
\end{center}
\caption{Frequency damping ratio $\zeta$ for the curves in Figures \ref{1-damped} (Model III) and \ref{A-damped} (Model I).}
\label{tab-damping-ratio}
\end{table}

\clearpage

\clearpage

\end{document}